**Dynamic heterogeneity of glass-forming liquids in the density scaling regime**

A. Grzybowski,* K. Koperwas, K. Kolodziejczyk, K. Grzybowska, and M. Paluch

*Institute of Physics, University of Silesia, Uniwersytecka 4, 40-007 Katowice, Poland*

* corresponding author's e-mail: andrzej.grzybowski@us.edu.pl

ABSTRACT

Recent analyses of high pressure measurement data suggest that the degree of the dynamic heterogeneity $\chi_4^{max}$ cannot be in general a single variable function of the structural relaxation time $\tau$. For a wide class of real and model supercooled liquids, the molecular dynamics of which obeys a density scaling law at least to a good approximation, we argue that the important relation between the length and time scales that characterize molecular motions near the glass transition is controlled by a density factor. If a power law density scaling is valid for both the structural relaxation times and the degrees of the dynamic heterogeneity we find that the factor is a density power, the exponent of which is a measure of the observed decoupling between $\tau$ and $\chi_4^{max}$. Then, the measure can be quantified by a difference between the power law density scaling exponents, which are usually different for $\tau$ and $\chi_4^{max}$.

A rapid slowdown in molecular dynamics of systems approaching the glass transition is still one of the most mysterious phenomena. In spite of intensive investigations in the past several decades, there is no complete theory of the glass transition and related phenomena, including the most pronounced one of them, i.e., the experimentally observed enormous increase in structural relaxation times τ or viscosities η of a liquid upon isobaric cooling within a relatively narrow temperature range or isothermal squeezing in the medium pressure limit. A prominent trend in the study of supercooled liquids invokes the assumption that their molecular motions have a heterogeneous character. This idea initiated by Adam and Gibbs[1] who suggested that interacting molecules of supercooled liquids form cooperatively rearranging regions, the size of which should increases upon approaching the glass transition, has been recently considerably developed by using a powerful formalism of a four-point time dependent correlation function $\chi_4(t)$,[2,3,4,5,6] suggested to study heterogeneous dynamics near the glass transition by Kirkpatrick and Thirumalai.[7] It has been achieved despite the fact that the function $\chi_4(t)$ for real glass formers is usually only estimated by performing a derivative analysis of a two-point correlation function $\Phi(t)$ like the stretched exponential function (i.e., $\Phi(t) = \exp[-(t/\tau)^\beta]$, where τ is the structural relaxation time and the stretching exponent, $0 < \beta \leq 1$), because the experimental measurements of $\chi_4(t)$ are very complex due to a need for detecting a nonlinear response of the examined sample. In theoretical and simulation studies, the dynamic susceptibility $\chi_4(t)$ that involves both temporal and spatial correlations is regarded as a precise measure of the dynamic heterogeneity, which quantifies a correlation volume, i.e., the volume that is characteristic for the correlated motions, and consequently the four-point time dependent correlation length, $\xi_4(t)$, by the relation, $\chi_4(t) \cong A(t)(\xi_4(t))^\psi$.[6] The peak of the dynamic susceptibility emerges at a timescale of the order of the relaxation time τ, which is typical for a given material in considered thermodynamic conditions. Depending on the assumed representation, the height of this peak, $\chi_4^{max}$, informs us on a typical correlation volume or a typical number of dynamically correlated molecules, $N_c$, at a given temperature $T$ and pressure $p$. In general, $N_c \sim \chi_4^{max}$,[2-6] and even an approximate correspondence between these quantities has been considered[3,8,9,10,11] in case of many systems, including bulk samples of simple van der Waals liquids. In the volumetric interpretation, one can expect[2] that a corresponding characteristic correlation length obeys the proportionality, $\chi_4^{max} \sim \left(\xi_4^{max}/a\right)^\psi$, where $a$ is a molecular size. For instance, according to the simplest approximation for bulk systems, $\psi=3$ and $a = \upsilon_m^{1/3}$, where $\upsilon_m$ is the molecular volume. Nevertheless, in our further discussion, we focus mainly on the measure $\chi_4^{max}$ of the characteristic correlation volume, over which structural relaxation processes are correlated,[4] to maintain a general level of our analysis, making it independent of particular relationships between $\chi_4^{max}$, $\xi_4^{max}$, and $N_c$.

The crucial achievement of the dynamic susceptibility formalism is the reliable determination of the temperature evolution of $\chi_4^{max}$ for systems approaching the glass transition. For both simulation and experimental data, it has been found that $\chi_4^{max}$ increases with decreasing temperature.[2-4] This result is really meaningful, because it confirms that the correlation length scale grows when a liquid upon isobaric cooling is approaching the glass transition. Very recently, we have shown that the isothermal squeezing of the liquid also increases its degree of the dynamic heterogeneity, $\chi_4^{max}$, although this effect is smaller than that caused by isobaric cooling.[11] A consequence of the latter is non-trivial, because it implies that the degree of the dynamic heterogeneity, $\chi_4^{max}$, decreases with increasing pressure at a constant structural relaxation time. What is more, this finding leads to a fundamental question *what is the proper relationship between the structural relaxation time and the degree of the* *dynamic heterogeneity?* In this paper, we show that such relations can be found for real and model supercooled liquids in the density scaling regime.

In recent years, much effort has been put into finding a proper relation between the characteristic scales of time and length for molecular dynamics of systems approaching the glass transition. For simple binary fluids, the relation $\tau \sim \xi^z$ has been discussed[12] for wavelengths longer than the correlation length $\xi$ liquid. Similarly, for kinetically constrained models, it has been suggested[13,14,15] that $\tau \sim \Xi^\zeta$, where $\Xi$ is the spacing between mobile elements. Thus, for simple models of supercooled liquids like the Kob-Andersen Lennard-Jones (KABLJ)[16] mixture, we can expect that $\ln\tau = f(\ln\xi)$. However, for real glass formers, different models of the temperature dependence of the structural relaxation time are considered, which are characterized by a general assumption that the energy barrier is proportional to $\xi^\psi$.[5,6,15,17,18,19,20,21] According to another quite common point of view, the structural relaxation time can simply be a function of a single variable such as the characteristic correlation volume or its corresponding length scale. For instance, comparing $\chi_4^{max}$ with $\tau$ for several glass formers at ambient pressure, Berthier *et al.*[3,4] has even observed a crossover from algebraic, $\chi_4 \sim \tau^z$, to logarithmic, $\chi_4 \sim \exp(\tau^\psi)$, growth of dynamic correlations with increasing $\tau$. However, from the mentioned high pressure study of $\chi_4^{max}$,[11] we can draw a conclusion that some previous results validated by using experimental data at ambient pressure cannot be held if we consider the complete thermodynamic space, because no single variable function, neither $f(\chi_4)$ nor $f(\xi_4)$, is able to describe the decrease in the degree of the dynamic heterogeneity with increasing pressure at $\tau = const$. Hence, the sought after relation between $\tau$ and $\chi_4^{max}$ should be more complex.

To remain a sufficiently unified level of theory in search of a proper dependence of $\tau$ on $\chi_4^{max}$, it is worth[22,23] invoking the density scaling of molecular dynamics near the glass transition. In general, this concept relies on many observations for important classes of glass formers, including mainly van der Waals supercooled liquids and polymer melts, which show that primary relaxation times (or viscosities) can be plotted onto one master curve versus a scaling variable $\Im(\rho)/T$.[24,25] In the most tempting case, $\Im(\rho) = \rho^\gamma$, where the scaling exponent γ is a material constant, which is suggested[26,27,28,29,30,31,32,33] to be straightforwardly related to the exponent, $m = 3\gamma$, of the repulsive inverse power law (IPL) term that constitutes the main part of an effective short range intermolecular potential which involves attractive interactions as a small background. The authors of previous combined studies of the power law density scaling and the dynamic heterogeneity have been trying to argue[8,34] that these quantities for both experimental and simulation data can be scaled with the same value of γ but in terms of different scaling functions $\tau = g(\rho^\gamma/T)$ and $\chi_4^{max} = h(\rho^\gamma/T)$. This point of view should be revisited in context of the change in $\chi_4^{max}(p)$ in isochronal conditions, which has been observed[10,11,35] for data obtained from different experimental techniques, because a composition of the function $g$ and the inverse function $h^{-1}$ results in a single variable function $\tau(\chi_4^{max})$. Thus, we postulate that $\chi_4^{max}$ can be scaled with the scaling exponent $\gamma_\chi$ which in general differs from γ that scales structural relaxation times. Within the framework of the power law density scaling idea, the scaling law with a material constant γ

$$\tau = g(\rho^\gamma/T) \qquad (1)$$

implies the following scaling condition[24,36]

$$\rho^\gamma/T = C_\tau \text{ at } \tau = const \qquad (2)$$

where $C_\tau$ depends only on the structural relaxation time. According to this manner, the scaling law for $\chi_4^{max}$ with a material constant $\gamma_\chi$

$$\chi_4^{max} = h(\rho^{\gamma_\chi}/T) \qquad (3)$$

leads to the condition

$$\rho^{\gamma_\chi}/T = C_\chi \text{ at } \chi_4^{\max} = const \qquad (4)$$

where $C_\chi$ depends only on the characteristic correlation volume. Then, from Eqs. (2) and (4) at $\tau = const$, we can establish a criterion for the combined power law density scaling of $\tau$ and $\chi_4^{\max}$,

$$\rho^{\Delta\gamma} = C_\tau/C_\chi \text{ with } \Delta\gamma = \gamma - \gamma_\chi, \qquad (5)$$

where $C_\tau = const$ and $C_\chi$ varies with changing density along an isochrone. Finding from Eq. (3) that $\rho^{\gamma_\chi}/T = h^{-1}(\chi_4^{\max})$ and exploiting Eqs. (1) and (5), we arrive at the non-trivial relation between the characteristic time scale and the measure of the characteristic correlation volume,

$$\tau = g\left(\rho^{\Delta\gamma} f(\chi_4^{\max})\right) \qquad (6)$$

where $\Delta\gamma = \gamma - \gamma_\chi$ and $f = h^{-1}$ is a single variable function of $\chi_4^{\max}$. This relation is a consequence of the power law density scaling of $\tau$ and $\chi_4^{\max}$ defined by Eqs. (1) and (3), respectively. In the general case of the density scaling, the single power of density in Eqs. (1) and (2) is not sufficient to achieve the scaling and we need to employ other usually more complex functions of density to do that.[37,38] It is reasonable to assume that the same problem concerns Eqs. (3) and (4) for the density scaling of $\chi_4^{\max}$ at the extremely high pressures. Thus, Eq. (5) should be rewritten as $\Re(\rho) = C_\tau/C_\chi$ with a density function, $\Re(\rho) = \Im_\tau(\rho)/\Im_\chi(\rho)$, where $\Im_\tau(\rho)/T$ and $\Im_\chi(\rho)/T$ determine the scaling variables for $\tau$ and $\chi_4^{\max}$, respectively. Then, Eq. (6) can be generalized as follows

$$\tau = g\left(\Re(\rho) f(\chi_4^{\max})\right) \qquad (7)$$

if $\tau$ and $\chi_4^{\max}$ obey the appropriate density scaling laws. Nevertheless, we limit the further tests to the case of the simple scaling variables, $\rho^\gamma/T$ and $\rho^{\gamma_\chi}/T$, because experimental data that comply with the power law density scaling rule are commonly accessible in contrast to those measured at sufficiently high pressures to reveal the more complex density scaling behavior.

The suggested relation can be verified by checking whether the observed decrease in $\chi_4^{max}(p)$ at $\tau = const$ can be reproduced by using Eq. (6). To perform the test we exploit experimental dielectric, volumetric, and heat capacity data for two typical van der Waals liquids, 1,1'-bis (p-methoxyphenyl) cyclohexane (BMPC)[39,40,11] and *o*-terphenyl (OTP),[41,42,43] previously considered by us[11] to discuss the effect of temperature and pressure changes on the degree of the dynamic heterogeneity, but herein we introduce another dielectric isobar of OTP measured[44] at ambient pressure in a considerably wider temperature range than that in ref. 41 and a recent parametrization[45] of PVT data for OTP in terms of the same equation of state[46] we have exploited to describe PVT data of BMPC in ref. 11. Nevertheless, values of $\chi_4^{max}(T,p)$ for BMPC and OTP are evaluated here in the same way we have used and described in detail in ref. 11, which is based on the estimator proposed by Berthier *et al.*,[2,3] $\chi_4^{max} \approx (\beta/e)^2 (\partial \ln \tau / \partial \ln T)^2 k_B / \Delta c_p$, where $\beta$ is the stretching exponent of the two-point correlation function $\Phi(t) = \exp[-(t/\tau)^\beta]$ and $\Delta c_p$ is the change in the heat capacity between liquid and glassy states. Then, the dependence $\chi_4^{max}(T,\rho)$ has been established with $\rho(T,p)$ found from the recently reported equation of state.[46]

It is already known[24,25] that structural relaxation times of BMPC and OTP obey the power law density scaling, which is illustrated in Figs. 1(a)-(b) and described by using the $\rho^\gamma/T$-scaling version of the Avramov entropic model,[47,48] $\tau = \tau_0 \exp[(A\rho^\gamma/T)^D]$, where $\tau_0$, $A$, $D$, $\gamma$ are fitting parameters. In this way, we choose an acknowledged representation[24] of the scaling function $g$ in Eq. (1). To verify whether the degree of the dynamic heterogeneity $\chi_4^{max}$ complies with a similar power law density scaling rule with a scaling exponent $\gamma_\chi$, we postulate that the scaling function $h$ in Eq. (3) can be given by, $\chi_4^{max} = (\chi_4^{max})_0 (A_\chi \rho^{\gamma_\chi}/T)^{D_\chi}$, where $(\chi_4^{max})_0$, $A_\chi$, $D_\chi$, $\gamma_\chi$ are fitting parameters. It is worth noting that the assumed form of the function $\chi_4^{max}(\rho^{\gamma_\chi}/T)$ is chosen to extend the logarithmic relation between $\tau$ and $\chi_4^{max}$ earlier suggested[3,4] by using experimental data at ambient pressure near the glass transition. We apply the function $\chi_4^{max}(\rho^{\gamma_\chi}/T)$ to fit the degrees of the dynamic heterogeneity evaluated at each $(T,p)$ at which dielectric measurements of BMPC and OTP have been carried out. As a result, we find that the quality of the fits (Figs. 1(c)-(d)) is even better (adjusted-$R^2$=0.99999) than that for the fits of structural relaxation times to the Avramov $\rho^\gamma/T$-scaling model (adjusted-$R^2$=0.99936 and 0.99831 for BMPC and OTP, respectively). Consequently, as far as we know for the first time, the scaling $\chi_4^{max}(\rho^{\gamma_\chi}/T)$ for real glass formers is achieved (see the insets in Figs. 1(c)-(d)) but with $\gamma_\chi<\gamma$. It means that the density factor $\rho^{\Delta\gamma}$ in Eq. (6) cannot be reduced to unity, because $\Delta\gamma>0$. It should be noted that if we evaluate $\chi_4^{max}$ and perform the scaling in terms of Eqs. (1) and (3) using $\tau$ in reduced units ($\tilde{\tau}=\tau\rho^{1/3}(k_BT/M)^{1/2}$ where $M$ is the average particle mass and $k_B$ is the Boltzmann constant), which ensure that *NVE* and *NVT* Newtonian dynamics are isomorph invariant and the scaling behavior is more clearly revealed,[31,32,49] the established values of $\Delta\gamma$ remain unchanged, i.e., 4.26±0.10 (4.30±0.10) for BMPC, and 1.04±0.05 (1.05±0.04) for OTP, where the values in brackets come from the analysis in the reduced units. This finding confirms our supposition that the relation between $\tau$ and $\chi_4^{max}$ cannot be in general described by a single variable function. Nevertheless, to complete our test, we need to check whether the found values of the scaling exponents $\gamma$ and $\gamma_\chi$ actually enable us to reproduce the decrease in $\chi_4^{max}(p)$ at $\tau=const$. For this purpose, we can apply the used scaling functions, $\tau(\rho^\gamma/T)$ and $\chi_4^{max}(\rho^{\gamma_\chi}/T)$, to construct the corresponding function $g$ in Eq. (6), and consequently, to formulate two equivalent equations,

$$\tau=\tau_0\exp\left[\left(\rho^{\Delta\gamma}(\chi_4^{max}/(\chi_4^{max})_0)^{1/D_\chi}A/A_\chi\right)^D\right] \text{ and } \chi_4^{max}=(\chi_4^{max})_0\left(\rho^{-\Delta\gamma}(\ln(\tau/\tau_0))^{1/D}A_\chi/A\right)^{D_\chi}.$$

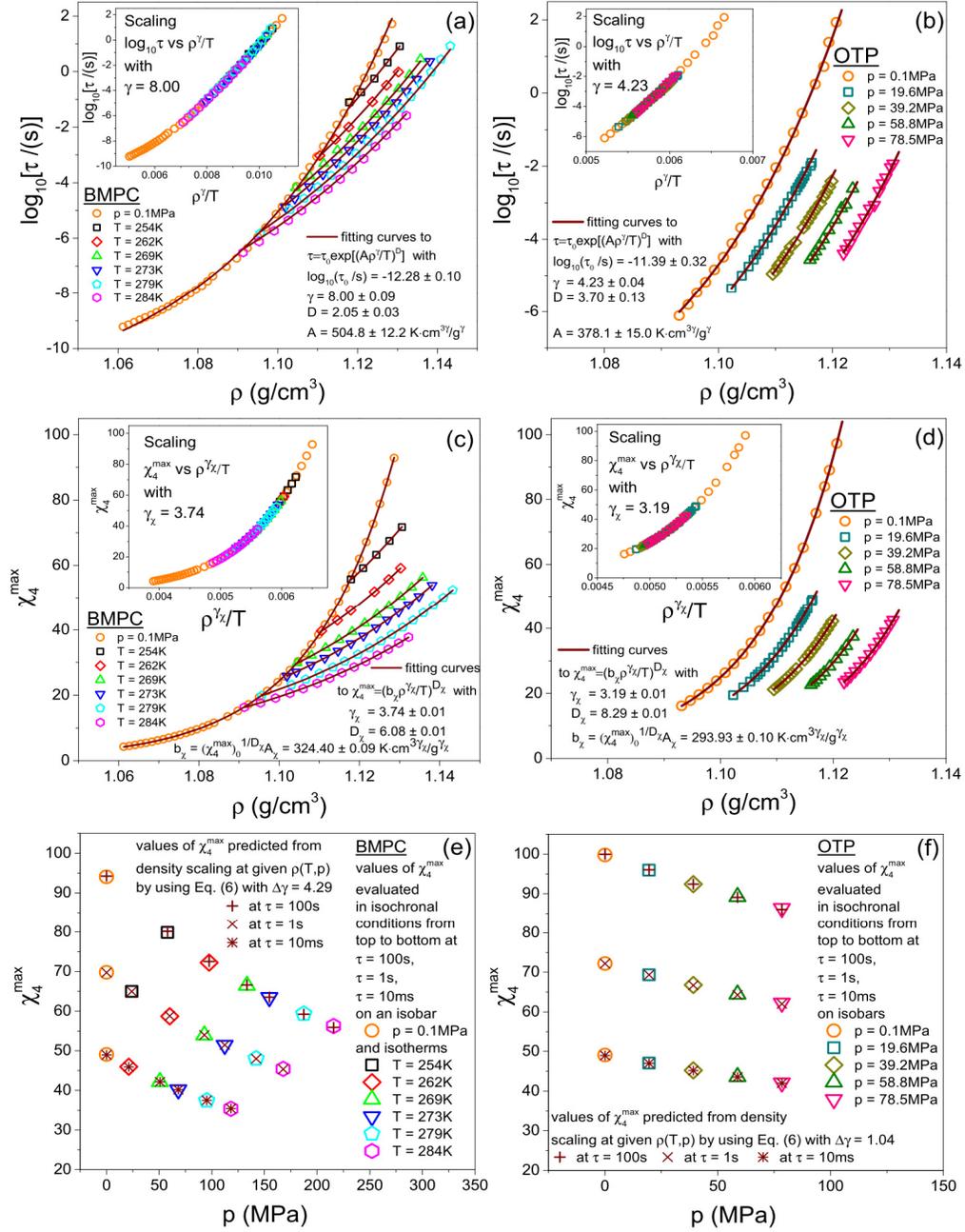

Fig. 1

The combined power law density scaling of $\tau$ and $\chi_4^{max}$ for real van der Waals supercooled liquids. (a)-(b) Plots of structural dielectric relaxation times vs density for BMPC and OTP. The insets in panels (a) and (b) present the scaling $\tau(\rho^\gamma/T)$ with the values of $\gamma$ found from the fitting shown in these panels. (c)-(d) Plots of the degree of the dynamic heterogeneity vs density for BMPC and OTP evaluated at each $(T,\rho)$ at which $\tau(T,\rho)$ is determined from experimental measurements. The insets in panels (c) and (d) present the scaling $\chi_4^{max}(\rho^{\gamma_\chi}/T)$ with the values of $\gamma_\chi$ found from the fitting shown in these panels. (e)-(f) Isochronal pressure dependences of the degree of the dynamic heterogeneity for BMPC and OTP. In panels (e) and (f), open symbols denote the evaluated values and the other symbols indicate values predicted from Eq. (6).

If we consider the latter equation with the values of its parameters taken from fitting $\tau$ and $\chi_4^{\max}$ to the scaling equations, $\tau = \tau(\rho^\gamma/T)$ and $\chi_4^{\max} = \chi_4^{\max}(\rho^{\gamma_\chi}/T)$, respectively, we should be able to reproduce the decreasing degree of the dynamic heterogeneity with increasing pressure in isochronal conditions. One can see in Figs. 1(e)-(f) that it can be indeed achieved for the examined van der Waals real supercooled liquids BMPC and OTP with $\Delta\gamma > 0$.

As a natural consequence, the next intriguing question arises as to whether the proposed formalism is valid for the simple models of supercooled liquids, which are widely exploited to investigate heterogeneous molecular dynamics by using direct measures of the dynamic heterogeneity that are naturally available in simulation studies, but are almost not applied to studies based on experimental data due to already mentioned large difficulties in measuring the four-point dynamic susceptibility. To answer the question we have performed[50] molecular dynamics (MD) simulations of a large system consisted of 8000 particles of the typical KABLJ liquid[16] in the relatively wide temperature-density range ($0.75 \leq T \leq 3.0$ and $1.2 \leq \rho \leq 1.6$ in LJ units). In our analysis, the structural relaxation times $\tau$ are calculated from simulation data in the usual manner,[32,33] i.e., by using[16] incoherent intermediate self-scattering functions ($\tau = t$ if $F_S(\mathbf{q},t) = e^{-1}$) determined at the wave vector $\mathbf{q}$ of the first peak of the structure factor for the particle species that dominates the binary content of the KABLJ system. Whereas the variance of the fluctuations of $F_s(\mathbf{q},t)$ suggested[51] to be a well-defined direct measure of the dynamic heterogeneity is used to establish the four-point dynamic susceptibility function $\chi_4(t)$, the maximum of which yields the value of $\chi_4^{\max}$.

It is already known that the structural relaxation times of the KABLJ liquid do not obey the power law density scaling in the exact sense,[28] but one can find an effective value of the scaling exponent $\gamma$ that enables the scaling in terms of the function $\tau(\rho^\gamma/T)$ to a good approximation at least in limited ranges of $T$ and $\rho$, which are, however, quite wide.[33]

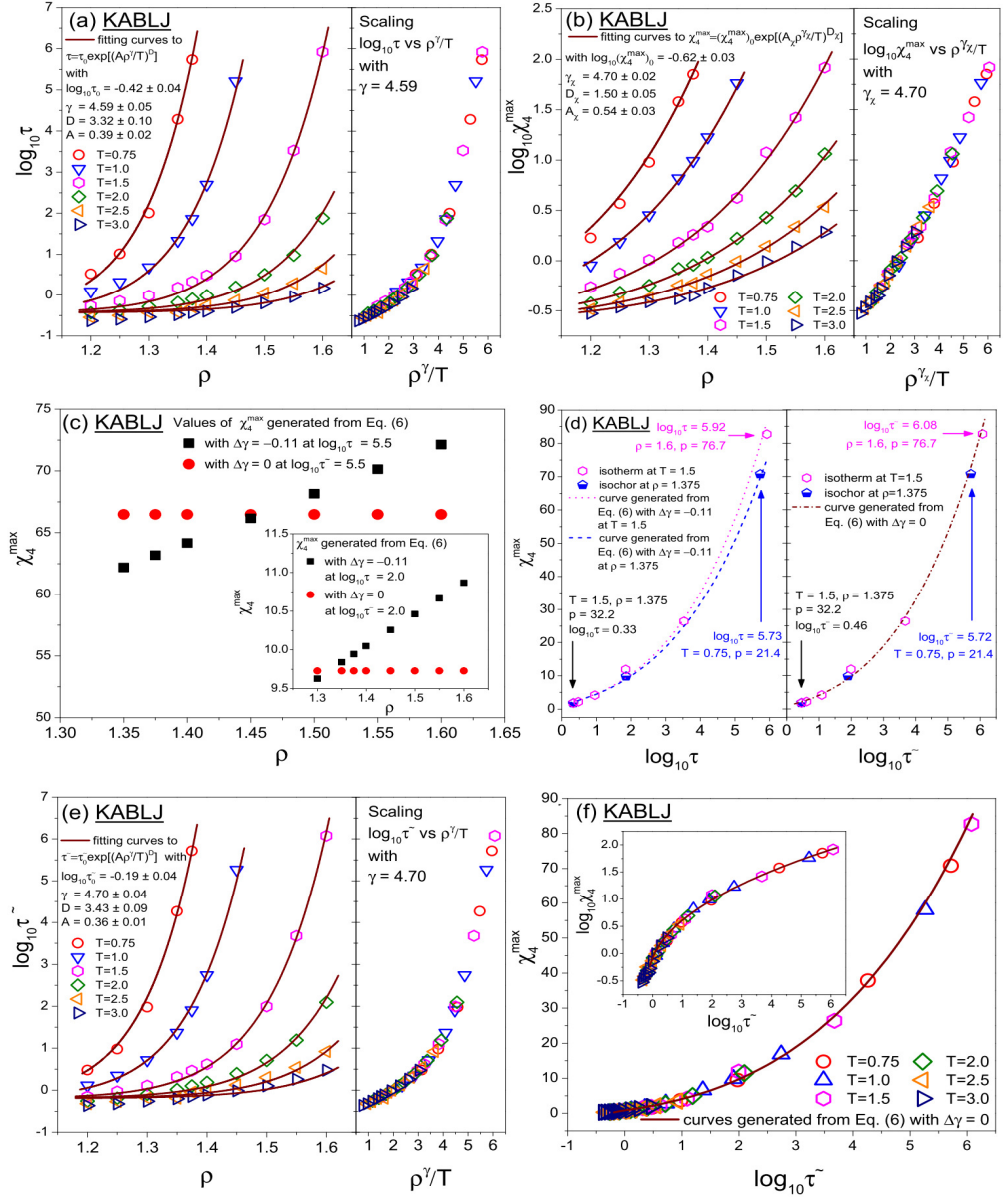

Fig. 2

The combined power law density scaling of $\tau$ and $\chi_4^{max}$ for the KABLJ model with 8000 particles. (a) Plot of the structural relaxation time in LJ units vs the particle number density in the left panel and the scaling $\tau(\rho^\gamma/T)$ in the right panel with the value of $\gamma$ found from the fitting shown in the left panel. (b) Plot of the degree of the dynamic heterogeneity vs the particle number density in the left panel and the scaling $\chi_4^{max}(\rho^{\gamma_\chi}/T)$ in the right panel with the values of $\gamma_\chi$ found from the fitting shown in the left panel. (c) Isochronal density dependences of the degree of the dynamic heterogeneity at long and (presented in the inset) short time scales expressed in LJ units ($\tau$) and the reduced units ($\tilde{\tau}$). (d) Isothermal and isochoric changes of the degree of the dynamic heterogeneity between two dynamic states characterized by $\log_{10}\tau \approx 0.3$ and $\log_{10}\tau \approx 5.9$ vs structural relaxation times expressed in LJ units (the left panel) and the reduced units (the right panel). (e) Plot of the structural relaxation time in the reduced units vs the particle number density in the left panel and the scaling $\tilde{\tau}(\rho^\gamma/T)$ in the right panel with the value of $\gamma$ found from the fitting shown in the left panel. (f) Plot of the degree of the dynamic heterogeneity vs the structural relaxation time in the reduced units. Its double logarithmic representation is shown in the inset.

By using the Avramov scaling model, we find here (Fig. 2(a)) the value γ=4.59±0.05 for the KABLJ system of 8000 particles, which enables us to scale τ in LJ units vs $\rho^\gamma/T$ to a good approximation. However, an attempt at applying the function previously used to scale the degree of the dynamic heterogeneity of real glass formers to determine a corresponding value $\gamma_\chi$ has ended with an insufficiently satisfactory fit of the dependence $\chi_4^{\max}(T,\rho)$. The result of the preliminary test can be understood in the context of earlier reports,[13,14,15] according to which one can rather expect as already mentioned that $\ln\tau = f(\ln\xi_4^{\max})$ for simple binary fluid models and kinetically constrained models, including the KABLJ liquid. In fact, a high quality of the fit has been found (Fig. 2(b)) for the T-ρ dependence of $\ln\chi_4^{\max}$ by means of the following scaling function, $\ln\chi_4^{\max} = \ln(\chi_4^{\max})_0 + (A_\chi \rho^{\gamma_\chi}/T)^{D_\chi}$, which morphologically corresponds to the Avramov scaling equation for the structural relaxation time, but it has in general different parameters. Consequently, for the KABLJ model and other simple binary fluid models, and presumably for kinetically constrained models, Eq. (6) can be interpreted as $\tau = g(\rho^{\Delta\gamma} f(\ln\chi_4^{\max}))$ if the power law density scaling is valid at least to a good approximation. By analogy, $\tau = g(\Re(\rho) f(\ln\chi_4^{\max}))$, is a version of Eq. (7), for simple models based on the Lennard-Jones potential and presumably for kinetically constrained models, which is the generalization about Eq. (6) not limited to a single power law density scaling. Invoking the Avramov scaling model, we can express Eq. (6) for the KABLJ model as follows $\tau = \tau_0 \exp\left[\left(\rho^{\Delta\gamma}\left(\ln(\chi_4^{\max}/(\chi_4^{\max})_0)\right)^{1/D_\chi} A/A_\chi\right)^D\right]$. Then, $\chi_4^{\max} = (\chi_4^{\max})_0 \exp\left[\left(\rho^{-\Delta\gamma}(\ln(\tau/\tau_0))^{1/D} A_\chi/A\right)^{D_\chi}\right]$. It is interesting that the scaling $\chi_4^{\max}(\rho^{\gamma_\chi}/T)$ for the KABLJ system examined in LJ units is achieved with the value $\gamma_\chi$=4.70±0.02, which is slightly greater than γ. Thus, in contrast to the tested real glass formers, $\Delta\gamma\mathrel{<}0$, which causes that $\chi_4^{\max}$ slightly increases with density (or pressure) at $\tau = const$ (see squares in Fig. 2(c)). This behavior resembles an isochronal increase in the fragility parameter $m_p = \partial \log_{10}(\tau)/\partial(T_g/T)|_p$ with density reported by Sastry[52] for the KABLJ model of 256 particles, which does not also match the usually decreasing[24,25,36] dependence of $m_p$ on pressure established from experimental data at the glass transition temperature $T_g$ defined at $\tau = const$, e.g., $\tau = 100s$. It should be emphasized that the suggested formalism that links the density scaling and the dynamic heterogeneity predicts an opposite effect of changes in temperature and density on the dynamic heterogeneity in the case of the negative value of $\Delta\gamma$ determined for the tested KABLJ liquid than that found for real glass formers. Very recently, we have evaluated[11] that the isochoric changes of $T$ more strongly exert on the dynamic heterogeneity than the isothermal changes of $\rho$ if we transform a real glass former from a state $(T,\rho)$ characterized by a structural relaxation time to another state at a different $\tau$. A thorough inspection of the degrees of the dynamic heterogeneity $\chi_4^{max}$ of the KABLJ liquid in isothermal and isochoric conditions as functions of the structural relaxation times $\tau$ confirms (the left panel of Fig. 2(d)) the predicted opposite behavior of the dynamic heterogeneity in this model on varying $\rho$ and $T$. However, the theory of isomorphs[49] recently formulated for viscous liquids clarifies that a hidden scale invariance[31,32] in such liquids can be revealed if the already mentioned reduced units are applied. Therefore, we investigate the molecular dynamics of the KABLJ liquid and its scaling properties also by using the reduced units (e.g., the structural relaxation time in the reduced units, $\tilde{\tau} = \tau\rho^{1/3}T^{1/2}$, if the Boltzmann constant and the particle mass are assumed to equal 1 in LJ units). As a result (Fig. 2(e)), we find a slightly better scaling $\tilde{\tau}(\rho^\gamma/T)$ than $\tau(\rho^\gamma/T)$, especially in the short-time limit, which is in accord with previously reported results[32,33] for the KABLJ system of 1000 particles. What is more, the value of $\gamma$ obtained from fitting $\tilde{\tau}$ to the Avramov scaling model matches the value $\gamma_\chi$ that enables us to scale $\chi_4^{max}$ vs $\rho^{\gamma_\chi}/T$.

This finding expected within the theory of isomorphs[49] and also established[34] for the KABLJ system of 1000 particles in a narrower density range is meaningful. Since $\Delta\gamma = 0$, Eq. (6) represents a single variable function of $\chi_4^{max}$ (see Fig. 2(f)), and consequently an isochronal state is also characterized by a constant value of $\chi_4^{max}$, which is independent of thermodynamic conditions and only depends on the structural relaxation time, e.g., $\chi_4^{max} = (\chi_4^{max})_0 \exp\left[\left(\left(\ln(\tilde{\tau}/\tilde{\tau}_0)\right)^{1/\tilde{D}} A_\chi / A\right)^{D_\chi}\right]$, which is exploited to generate points at $\tilde{\tau} = const$ (circles in Fig. 2(c)). Moreover, in this case, the transition from one isochronal state to another one at different $\tilde{\tau}$ should be independent of the thermodynamic path toured between these two dynamic states. In the right panel of Fig. 2(d), one can see that the isochoric and isothermal paths indeed superimpose. It means that the isochoric changes of $T$ and the isothermal changes of $\rho$ equivalently affect the dynamic heterogeneity in the KABLJ model examined in the reduced units.

From our studies of real supercooled liquids measured at ambient and high pressures and the KABLJ model in the relatively wide temperature-density range, we can conclude that the degree of the dynamic heterogeneity $\chi_4^{max}$ cannot be in general a single variable function of the structural relaxation time $\tau$, but it requires an additional factor. If molecular dynamics obeys the density scaling law at least to a good approximation, this factor depends only on density. In the case of the density scaling described by the power laws, $\tau(\rho^\gamma/T)$ and $\chi_4^{max}(\rho^{\gamma_\chi}/T)$, we show that the proper relation between $\tau$ and $\chi_4^{max}$ requires the power density factor $\rho^{\Delta\gamma}$. Its exponent, $\Delta\gamma = \gamma - \gamma_\chi$, can be regarded as a measure of the observed decoupling between $\tau$ and $\chi_4^{max}$, which can be, however, eliminated in some systems as it is achieved in the KABLJ model by using the reduced units that ensure the isomorph invariance of the molecular dynamics.

The observed decoupling between $\tau$ and $\chi_4^{max}$ of real glass formers changes our understanding of molecular dynamics near the glass transition, because it suggests that although the time and length scales of molecular dynamics both increase on approaching the glass transition in isobaric or isothermal conditions, the time and length representations are not equivalent. It means that a given time scale ($\tau = const$) involves various length scales (related to $\chi_4^{max}$) depending on thermodynamic conditions. This intriguing behavior seems to be an intrinsic feature of real glass formers, which should be considered in further theoretical and experimental investigations.

**Acknowledgements**

The authors gratefully acknowledge the financial support from the Polish National Science Centre within the program MAESTRO 2. K.K. is deeply thankful for the stipend received within the project "DoktoRIS - the stipend program for the innovative Silesia", which is co-financed by the EU European Social Fund.

[24] Floudas, G., Paluch, M., Grzybowski, A. & Ngai, K. *Molecular Dynamics of Glass-Forming Systems: Effects of Pressure*, Ch. 2, Series: *Advances in Dielectrics*, Series Editor: Friedrich Kremer, (Springer-Verlag, Berlin, Heidelberg, 2011).

[25] Roland, C. M., Hensel-Bielowka, S., Paluch, M. & Casalini, R. Supercooled dynamics of glass-forming liquids and polymers under hydrostatic pressure. *Rep. Prog. Phys.* **68**, 1405 - 1478 (2005).

[26] Alba-Simionesco, C. & Tarjus, G. Temperature versus density effects in glassforming liquids and polymers: A scaling hypothesis and its consequences. *J. Non-Cryst. Solids* **352**, 4888-4894 (2006).

[27] Pedersen, U. R., Bailey, N. P., Schrøder, T. B. & Dyre, J. C. Strong pressure-energy correlations in van der Waals liquids. *Phys. Rev. Lett* **100**, 015701-1 - 015701-4 (2008).

[28] Bailey, N. P., Pedersen, U. R., Gnan, N., Schrøder, T. B. & Dyre, J. C. Pressure-energy correlations in liquids. I. Results from computer simulations. *J. Chem. Phys.* **129**, 184507-1 - 184507-13 (2008).

[29] Coslovich, D. & Roland, C. M. Thermodynamic scaling of diffusion in supercooled Lennard-Jones liquids. *J. Phys. Chem. B* **112**, 1329-1332 (2008).

[30] Coslovich, D. & Roland, C. M. Pressure-energy correlations and thermodynamic scaling in viscous Lennard-Jones liquids. *J. Chem. Phys.* **130**, 014508-1 - 014508-5 (2009).

[31] Schrøder, T. B., Pedersen, U. R., Bailey, N. P., Toxvaerd, S. & Dyre, J. C. Hidden scale invariance in molecular van der Waals liquids: A simulation study. *Phys. Rev. E* **80**, 041502-1 - 041502-6 (2009).

[32] Pedersen, U. R., Schrøder, T. B. & Dyre, J. C. Repulsive reference potential reproducing the dynamics of a liquid with attractions. *Phys. Rev. Lett.* **105**, 157801-1 - 157801-4 (2010).

[33] Grzybowski, A., Koperwas, K. & Paluch, M. Scaling of volumetric data in model systems based on the Lennard-Jones potential. *Phys. Rev. E* **86**, 031501-1 - 031501-9 (2012).
18